\documentclass{elsart}

\usepackage{graphicx}

\usepackage{amssymb}

\begin{document}

\begin{frontmatter}



\title{A Study of the Water Cherenkov Calorimeter}

\author[ihep,ustc]{M.-J. Chen\corauthref{cor1}}
\corauth[cor1]{E-mail address: mjchen@mail.ihep.ac.cn}
\author[ihep]{Y.-F. Wang}
\author[ihep]{J.-T. He}
\author[ihep]{M.-L. Yu\thanksref{hnu}}
\address[ihep]{Institute of High Energy Physics, Beijing 100049, China}
\address[ustc]{University of Science and Technology of China, Hefei 230026,
 China}
\thanks[hnu]{Present address: HuaZhong Normal University, Wuhan 430079, China}

\begin{abstract}
The novel idea of water Cherenkov calorimeter made of water tanks as
the next generation neutrino detector for $\nu$ factories and $\nu$
beams is investigated. A water tank prototype with a dimension of
$1\times 1\times 13m^3$ is constructed, its performance is studied
and compared with a GEANT4 based Monte Carlo simulation. By using
measured parameters of the water tank, including the light
collection efficiency, attenuation length, angular dependent
response etc, a detailed Monte Carlo simulation demonstrates that
the detector performance is excellent for identifying neutrino
charged current events while rejecting neutral current and
wrong-flavor backgrounds.
\end{abstract}

\end{frontmatter}

\section{Introduction}

Neutrino factories and conventional beams have been discussed
extensively~\cite{neufac,chen01} as the facility of neutrino physics
for the next decade. The main physics objectives include the
measurement of $\sin\theta_{13}$, $\Delta m^2_{13}$, the leptonic CP
phase $\delta$ and the sign of $\Delta m^2_{23}$. All these
quantities can be obtained through the disappearance probability
$P(\nu_{\mu} \to \nu_{\mu})$ and the appearance probability
$P({\nu_{\mu}}({\nu_e}) \to {\nu_e}({\nu_{\mu}}))$ and
$P(\bar{\nu_{\mu}}(\bar{\nu_e}) \to \bar{\nu_e}(\bar{\nu_{\mu}}))$.
To measure these quantities, a detector should: 1) be able to
identify leptons: $e, \mu$ and if possible $\tau$; 2) have good
pattern recognition capabilities for background rejection; 3)have
good energy resolution for event selection and to determine
$P_{\alpha \to \beta}(E)$; 4) be able to measure the charge for
$\mu^{\pm}$ in the case of $\nu$ factories; and 5) be able to have a
large mass(100-1000kt) at an affordable price.

Water is one of the most economic materials for large scale neutrino
detectors. Water Cherenkov ring image detectors have been
successfully employed in large scale experiments such as
Super-Kamiokande~\cite{superk}, MiniBooNE~\cite{BNE} and
IMB~\cite{IMB}, etc. However such kind of detectors are not suitable
for neutrinos with an energy more than $\sim 4GeV$ due to
complications of showers, therefore not the choice for very long
baseline neutrino oscillation experiments. The water Cherenkov
calorimeter made up by a matrix of water tanks, was
proposed~\cite{wcc0} for the long baseline neutrino oscillation
experiments.

\begin{figure}[htbp]
\begin{center}
\includegraphics*[width=0.7\textwidth]{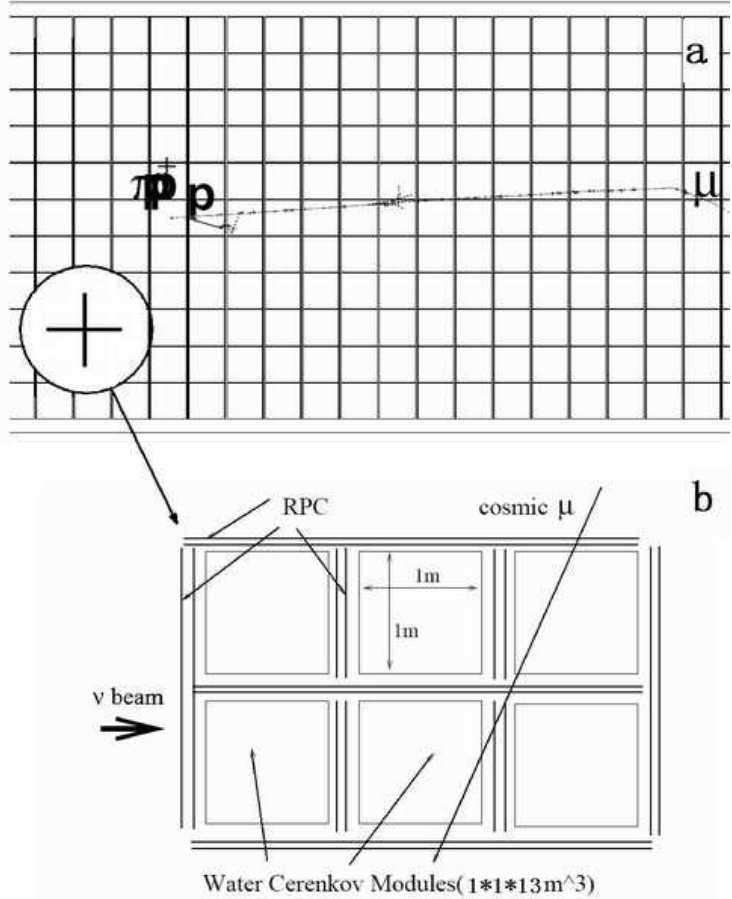}
\begin{minipage}[c]{10cm}
\caption{Schematics of the water Cherenkov calorimeter and a typical
$\nu_\mu$ CC events in the detector.} \label{fig:event}
\end{minipage}
\end{center}
\end{figure}

A water Cherenkov Calorimeter with a modular structure is shown in
Fig.\ref{fig:event}. Each tank has dimensions of about $1\times
1\times 13 m^3$, corresponding to 2.77X$_0$ and 1.5$\lambda_0$ in
its transverse dimension. The water tank is made of PVC with
reflective inner lining. Photons are collected at both end of the
tank, hence significantly reducing the photon collection area and
the cost. Cherenkov light produced by charged particles in the water
tank is estimated to be sufficient for energy resolution, and
position is determined by the arrival time of photons to the
phototubes at both ends of the tank. The directional information is
obtained by the reconstruction of the event thrust. The event
pattern in energy and space in the water tank matrix can be used to
identify neutrinos undergoing charge current(CC) interactions, as
shown in Fig. \ref{fig:event}.a which is a typical $\nu_{\mu}$ CC
event.


The water Cherenkov calorimeter is similar in a sense to the crystal
calorimeter at accelerator experiments. It is a cheap solution for
the long baseline neutrino oscillation experiments at a scale of
100-1000kt, and also applicable to cosmic-ray physics and
astrophysics. Reports about these kind of applications can also be
found in Ref.~\cite{tank}. In this paper, we report our study with a
prototype and a Monte Carlo simulation using measured parameters of
the water tank performance for the identification of neutrino CC
events.

\section{Water Tank Prototype}

\begin{figure}[htbp]
\begin{center}
\includegraphics[width=.95\textwidth]{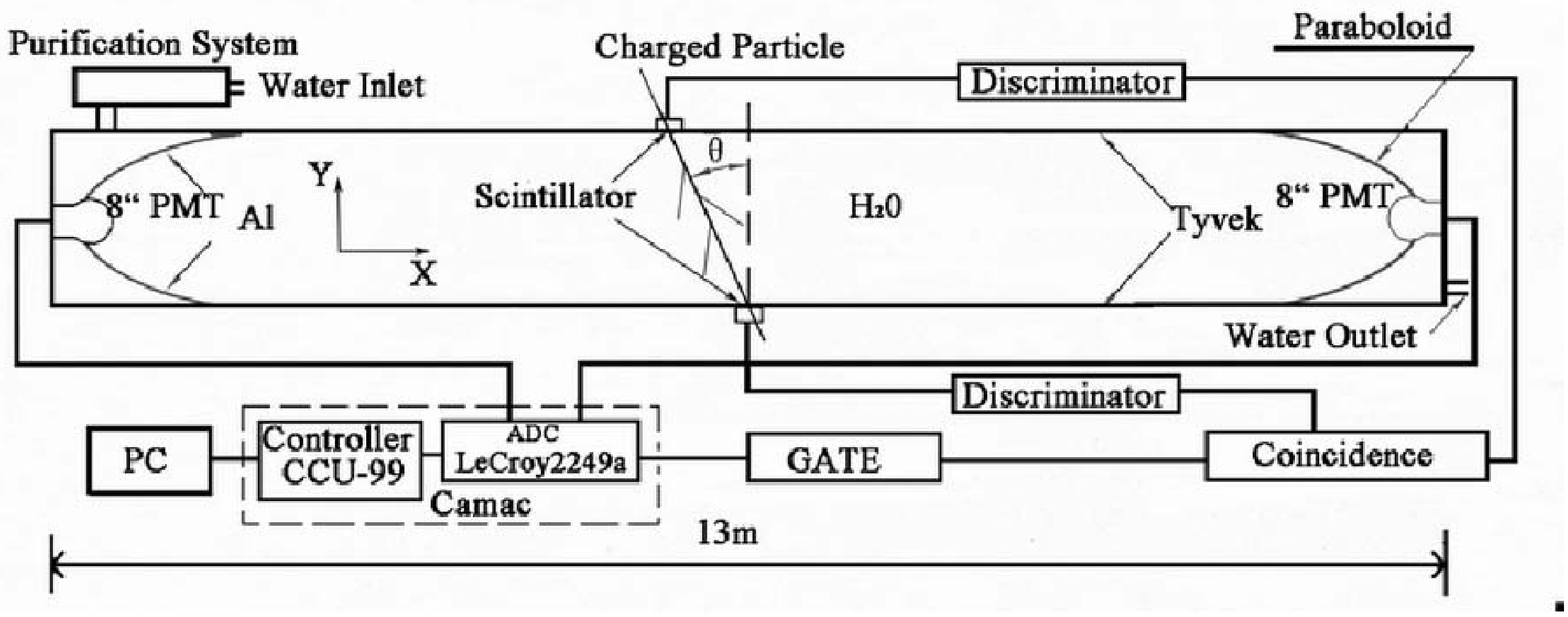}
\begin{minipage}[c]{10cm}
\caption{ Schematics of a water tank. The middle point of the tank
is set as the origin of the coordinate system. \label{setup}}
\end{minipage}
\end{center}
\end{figure}

A water tank prototype~\cite{wcc2} made of PVC with the dimension of
$1\times 1\times 13m^3$ is built as shown in Fig. \ref{setup}. The
inner wall of the tank is covered by the Tyvek film 1070D from
DuPont. At each end of the tank there is a Winston
cone~\cite{winston} which can collect parallel light at its focal
points, where an 8-inch photomultiplier is installed.  The Winston
cone is again made of PVC, covered by the aluminium film with
protective coating. Cherenkov light produced by through-going
charged particles are reflected by the Tyvek and the Al film and
collected by the photomultiplier at the focus of the Winston cone.
At the top of the tank there is an air gap(about 1cm) above the
water level which serves as a total reflector for photons with
certain incident angles.

\begin{figure}[htbp]
\begin{center}
\includegraphics[width=0.65\textwidth]{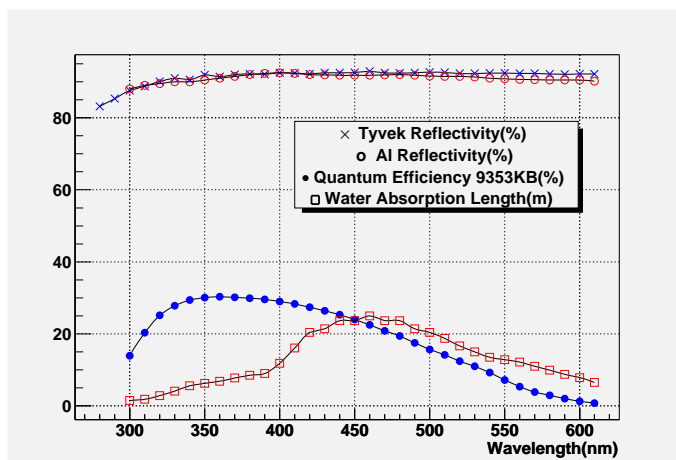}\\
\begin{minipage}[c]{10cm}
\caption{The Water absorption length by adjusting that from Auger
experiment, the quantum efficiency of the PMT 9350KB from EMI, and
the measured reflectivity of Tyvek and Al as a function of
wavelength. \label{RQA}}
\end{minipage}
\end{center}
\end{figure}

Tyvek is a diffuse reflector with a very high reflectivity which is
measured in the frequency range of visible light as shown in Fig.
\ref{RQA}. Although it is naively believed that mirror reflector
such as the Al film has a better light collection for such a long
optical module, our simulation shows that their performances are
actually very similar~\cite{wcc1}. The dominant factor is the bulk
reflectivity. The good mechanical and chemical properties of the
Tyvek film lead us to use it in order to have an easy handling and
less aging effect in the deionized water. Tyvek as a reflector in
water has been used by many experiments, including Super-Kamiokande,
KamLAND, and Auger experiments~\cite{augertyvek}.

Since the Winston cone needs a mirror reflection to collect light, a
selected Al film is used. Al film has a very high
reflectivity$(~98\%)$ in theory, but is easy to be oxidized in water
and loss its reflectivity. A protective coating is hence needed and
the reflectivity is measured to be typically 90\%, as shown in Fig.
\ref{RQA}.

In order to have a good water transparency, the clean de-ionized
water with a resistance of more than $10M \Omega \cdot cm$ is used.
The water is again purified by a simple system with a $0.1\mu m$
filter, which can increase the transparency by a factor of two. The
water absorption length as a function of wavelength used in Monte
Carlo simulation is obtained by scaling down the curve from the
Auger experiment~\cite{augertank} based on our experimental data, as
shown in Fig. \ref{RQA}. The phototube used is 9350KB from EMI, and
its quantum efficiency~\cite{pmt} is shown in Fig. \ref{RQA}.

Cosmic-muons, triggered by two scintillator counters at the top and
the bottom of the tank, are the primary charged particles which
generate Cherenkov photons. The muon flux at the sea level is about
$1.8\times 10^{-2}/cm^2 \cdot s$, and the area of scintillation
counters is $20cm \times 44cm$, hence it takes typically 10 hours to
accumulate one spectrum. Such a small trigger counter is selected to
control the error due to the incident position, angle and the pass
length of muons. A displacement of one of the two trigger counters
in x direction(see Fig. \ref{setup}) can define the incident angle
of the muons. In addition, between two cosmic data taking runs a
calibration run with the trigger counters at x=0.5m was taken to
monitor the water quality.

The setup includes also a C205 ADC from CAEN to measure the charge
of muon and the single photoelectron for calibration, a N844
discriminator from CAEN to generate trigger signals and the gate
signal for ADC.

\subsection{Monte Carlo Simulation}
Since Geant4 provides quite powerful tools for describing the
detector and the relevant physics with a fairly flexible optical
model inspired by the work of Nayer {\it et al.}~\cite{nayer}, a
GEANT4 based Monte Carlo simulation program of this water tank
prototype has been developed.

The so called UNIFIED model, which accommodates the principal
features of both physical and geometrical optical models of surface
reflection over a wide range of surface roughness and wavelengths,
is used in our simulation. The implementation of UNIFIED model
requires the specification of seven free
parameters~\cite{geant4surface}: $n_1$ is the index of refraction of
the incident medium; $n_2$ the index of refraction of the
transmission medium; $C_{sl}$ the specular lobe constant; $C_{ss}$
the specular spike constant; $C_{bs}$ the backscatter constant;
$C_{dl}$ the diffuse lobe constant; and $\sigma_{\alpha}$ the RMS of
the angle $\alpha$ between the average surface normal and microfacet
normals. Since the construction of the water tank is similar in many
ways to that of the Auger detector, some values of the parameters in
our simulation programs are selected to be very similar
~\cite{augerprogram}. We set $C_{sl}=0.2$ with the corresponding
$\sigma_{\alpha}=0.2$, $C_{ss}=C_{bs}=0.$, and $C_{dl}=0.8$. $n_1$
and $n_2$ are from our experimental measurements. More details about
optical models and its parameters have been discussed in
Ref.~\cite{wcc1,auger3}.

\subsection{Experimental study of the Water Tank Prototype}

\subsubsection{PMT's Single-photoelectron Spectrum}

\begin{figure}[htbp]
\begin{center}
\includegraphics[width=0.8\textwidth]{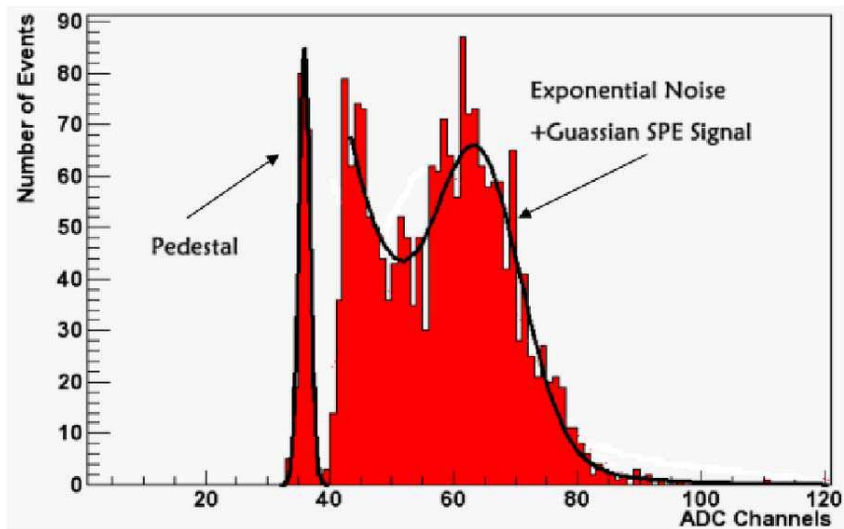}\\
\begin{minipage}[c]{10cm}
\caption{Single-photon spectrum of the PMT 9350KB
 at a high voltage of 1550V. \label{PMT}}
\end{minipage}
\end{center}
\end{figure}

Single photoelectron spectrum(SPE) is measured before each run in
order to calibrate the system since signal amplitudes normalized to
that of SPE provide a unique measure of light collected by
photomultipliers. SPE can be measured in many ways, one of which is
the so called "thermal noise" method. In total darkness, a
photomultiplier can still generate pulses due to thermal emission of
single electron by photocathode, equivalent to the charge spectrum
of single photoelectron. Thermal emission of electrons by dynodes
constitutes the noise below the SPE peak. A SPE spectrum of the PMT
9350KB, applying a high voltage of 1550V at the room
temperature(about $15^o$C), is measured as shown in Fig. \ref{PMT}.
Since the ADC used is only 12 bit, the working voltage(1550V) of the
PMT is selected to avoid saturation of ADC for cosmic-muons at all
positions along the water tank. The SPE spectrum is obtained by a
self-trigger with a threshold of 2mV and a gate width of 100ns. The
first peak corresponds to the pedestal, the second peak comes from
the dynode noise above the 2mV threshold, and the last peak is from
SPE, whose position will be used as the normalization to count
number of photoelectrons.

The effective attenuation length of water had worsened over time,
hence calibration runs were taken before each measurement, and all
data points in the late sections are after correction.

\subsubsection{Position Dependent Response of the Water Tank}

Light collected for cosmic-muons is a function of distance from the
incident point of the muon to the phototube, since the water
transparency and reflectivity of the Tyvek film is not perfect. Such
a position dependent response of the tank is critical to its energy
resolution and pattern recognition capability. Typically it is
characterized by an exponential behavior of $e^{-x/\lambda}$, where
x is the distance of the muon event to the phototube and $\lambda$
is the characteristic parameter, often called
 "effective attenuation length".

\begin{figure}[htbp]
\begin{center}
\includegraphics[width=0.60\textwidth]{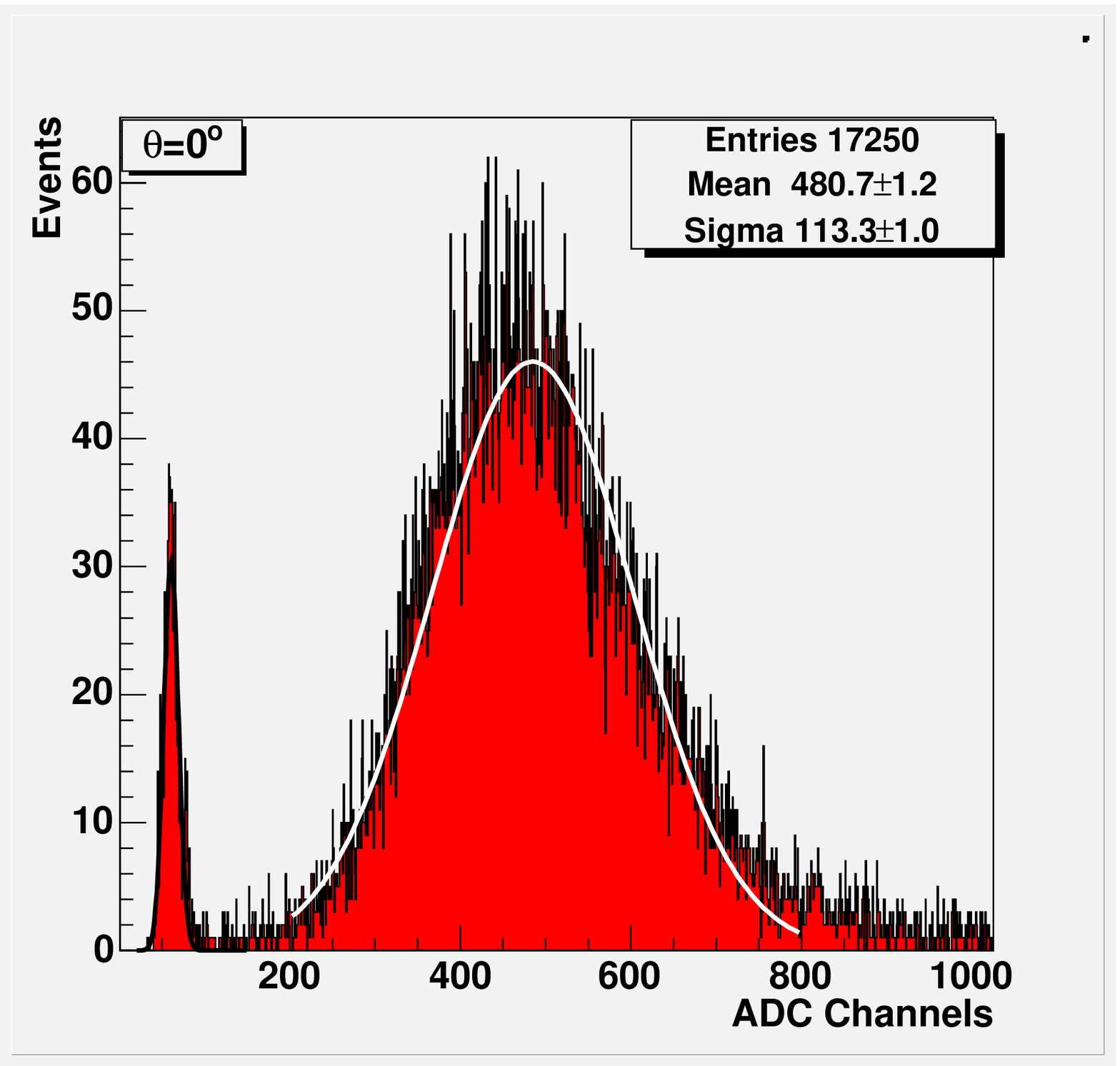}\\
\begin{minipage}[c]{10cm}
\caption{Charge spectrum collected at x=0.5m with an incident angle
of $0^o$. \label{data}}
\end{minipage}
\includegraphics[width=0.70\textwidth]{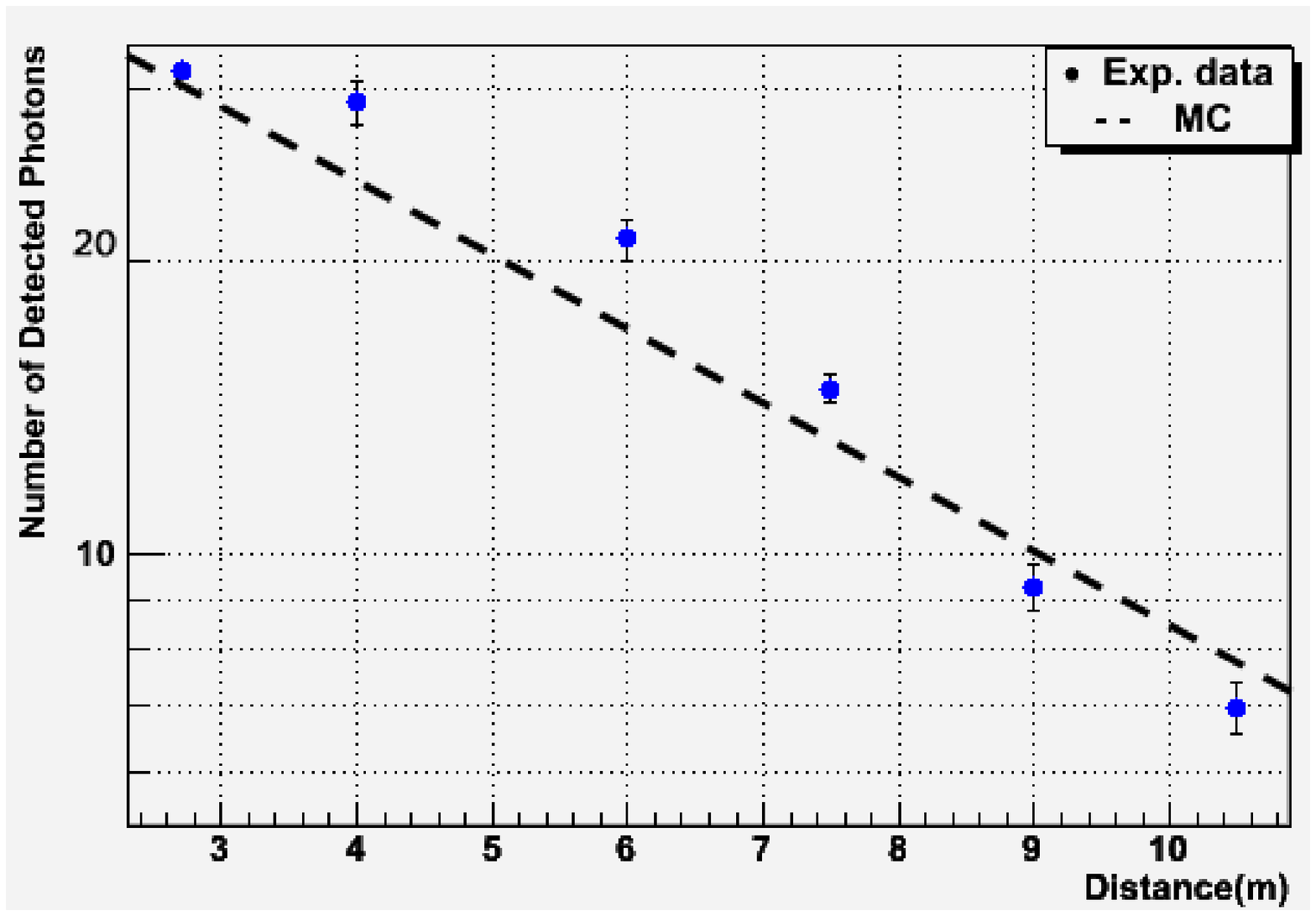}\\
\begin{minipage}[c]{10cm}
\caption{Position dependent response of the water tank to
cosmic-muons. X is the distance from trigger counters to the PMT at
right. The line represent the Monte Carlo prediction with an
effective attenuation length of 5.79m. The measured effective
attenuation length of the water tank is (5.74$\pm$0.29)$m$.
\label{wabs}}
\end{minipage}
\end{center}
\end{figure}

The characteristic parameter $\lambda$ depends on the water
transparency, the reflectivity of the Tyvek film, and the geometry
of the tank. Fig. \ref{data} shows the charge spectrum collected at
x=0.5m with an incident angle of $0^o$. Using the trigger
scintillation counters to define the muon incident location, keeping
the y coordinate constant as indicated in Fig. \ref{setup}, the
total light collected as a function of x at several locations is
obtained as shown in Fig. \ref{wabs}. An exponential fit yields the
measured effective attenuation length of the water tank of $(5.74
\pm 0.29)m$. The line represents the Monte Carlo prediction by
adjusting the water absorption length as shown in Fig. \ref{RQA},
until the effective attenuation length is in agreement with that of
the measurement. As to be discussed later, this tuning is justified
by the agreement between data and Monte Carlo prediction for both
the effective attenuation length and the angular dependent response.

It can be seen from Fig. \ref{wabs} that, for a through-going muons
entering the center of the tank, a total of $\sim$20 photoelectrons
by each PMT will be collected, corresponding to a statistical
fluctuation of about 7\%/$\sqrt{E}$. Based on the Monte Carlo
simulation, the number of photons at various stages of the photon
transport in the water tank is listed in table \ref{photons}. From
the table, about 74\% of light lost due to the Tyvek reflection and
water absorption. The Winston cone has a collection efficiency of
3.1\%, same as the ratio of PMT surface area to that of the water
tank cross section. It means that the Winston cone did not improve
the light collection efficiency, but the uniformity of the light
collection.

There are several ways to improve the light collection of the water
tank: a) The water absorption length can be improved with a more
sophisticated purification system. In fact the Super-Kamiokande
experiment reached an absorption length of about 90m~\cite{SKabs}, a
factor of 3 better than what was reached here; b) The reflectivity
of the inner liner can be improved by using newly developed plastic
reflectors, VM2000 or ESR from 3M Co.~\cite{3Mcor}. They have a
reflectivity better than 99\%, which can increase the total light
collected by more than 50\%. In total, it is possible to increase
the light collection by a factor of two, corresponding to a
statistical fluctuation of about
5\%/$\sqrt{E}$ for each tank.\\

\begin{table}[htbp]
\begin{center}
\begin{minipage}[c]{10cm}
\caption{Number of photons at various stages of the photon transport
in the water tank from Monte Carlo simulation. \label{photons}}
\end{minipage}
\begin{tabular}{l|l}\hline No. of Cherenkov photons produced &  35157$\pm$179  \\ \hline
No. of photons entering Winston cones & 9274$\pm$76 \\  \hline No.
of Photons hitting the glass surface of two PMTs  & 288$\pm$18
\\ \hline No. of photoelectrons collected by two PMTs &
42$\pm$3\\ \hline
\end{tabular}
\end{center}\
\end{table}

\subsubsection{Angular Dependent Response of the water tank}

\begin{figure}[htbp]
\begin{center}
\includegraphics[width=0.65\textwidth]{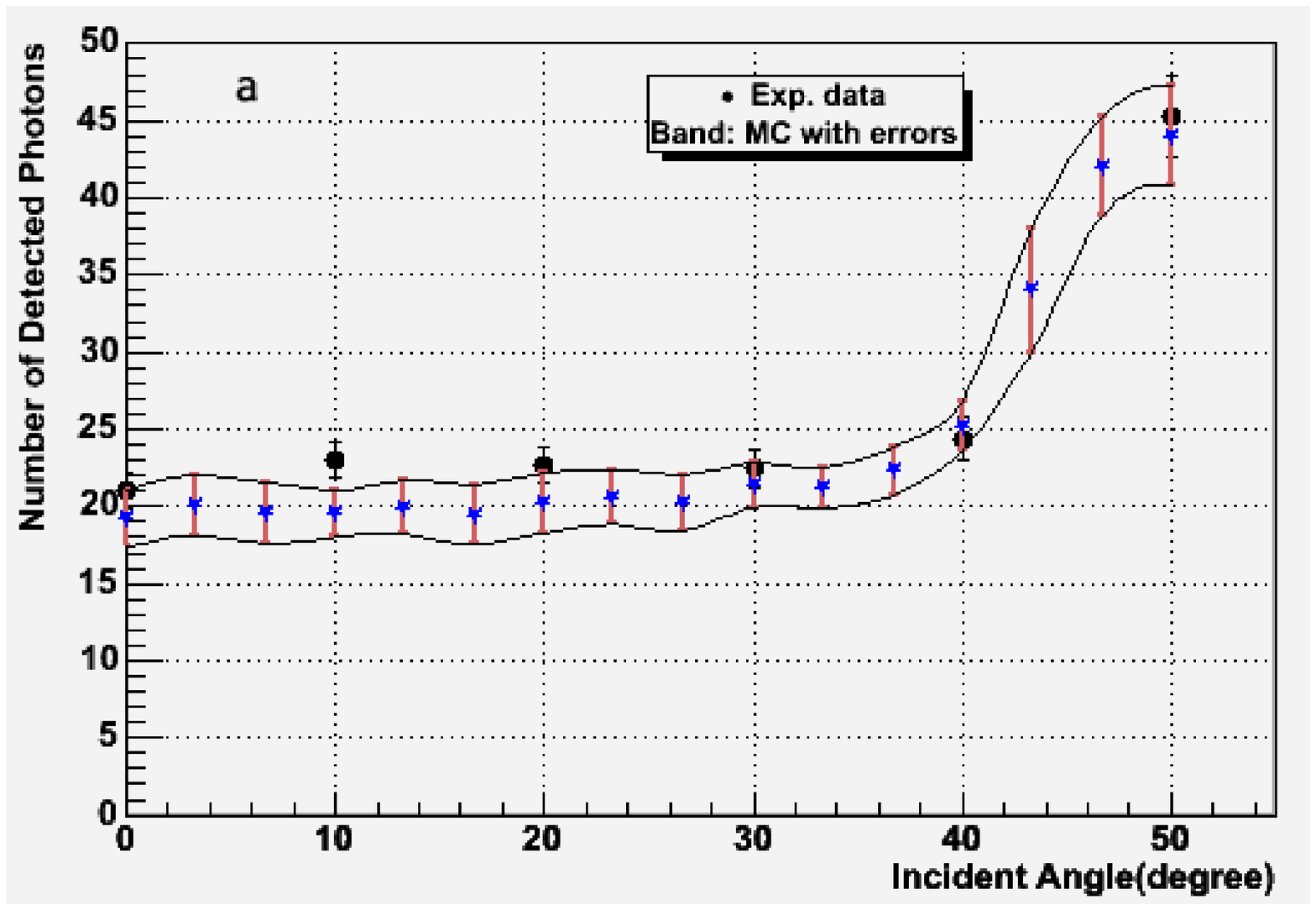}\\
\includegraphics[width=0.65\textwidth]{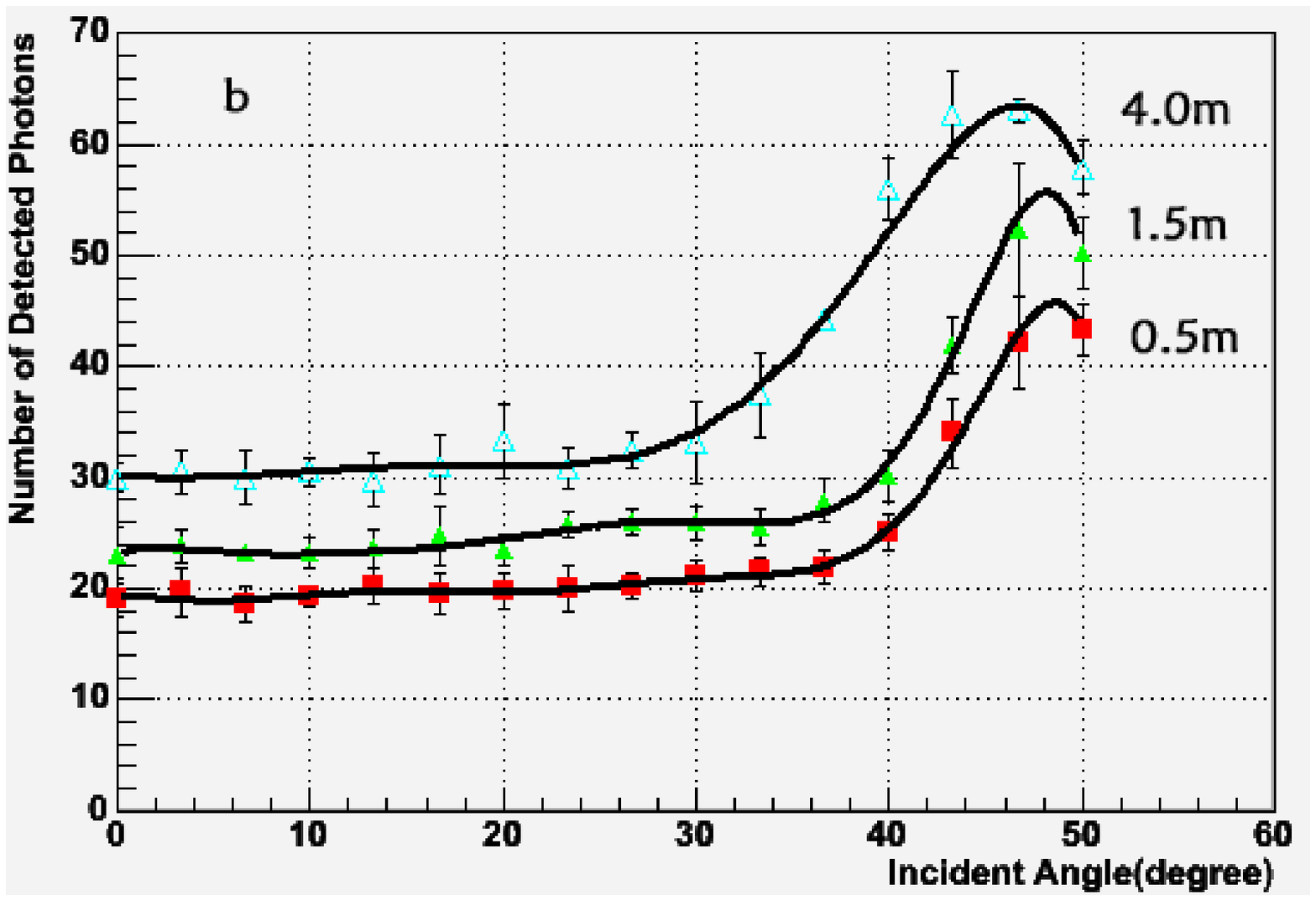}\\
\begin{minipage}[c]{10cm}
\caption{a) Measured angular dependent response of the water tank
together with Monte Carlo prediction. The band indicate the
statistical error of the Monte Carlo prediction. The track length of
all the data points are normalized to 1 meter; b) Monte Carlo
results of the angular response as a function of distance from the
incident point to the phototube. \label{angle}}
\end{minipage}
\end{center}
\end{figure}

Since Cherenkov light produced is not isotropic, and its direction
is correlated to that of the incident charged particles, the total
light collected by phototubes at each end of the water tank is also
correlated to the incident angle of the particles. By using trigger
counters to define the angle as shown in Fig. \ref{setup}, response
of the water tank to through-going charged muons with incident
angles varied from $0^o$ to $50^o$ are measured. The bottom trigger
scintillator is fixed at x=0.5m, and the top trigger scintillator is
moved along the -x direction. After normalizing the track length to
1m, results are shown in Fig. \ref{angle}a together with predictions
from the Monte Carlo simulation. Since the only free parameter to be
tuned in the Monte Carlo prediction is the overall scaling of the
water absorption length as discussed before, the good agreement
between data and Monte Carlo simulation for both effective
attenuation length and the angular dependent response shows that the
optical behavior of the water tank is largely understood.

As can be seen from Fig. \ref{angle}a, the number of photoelectrons
is approximate constant for incident angles less then 3$0^o$. This
is confirmed by the Monte Carlo simulation, and true at almost all
locations of the tank, as shown in Fig. \ref{angle}b. This is
significant since during the event reconstruction, this factor can
be ignored and the energy resolution of neutrino event can be
maintained at a reasonably good level.

\section{the Water Cherenkov Calorimeter for $\nu$ Detection}

Water Cherenkov calorimeter for the neutrino detection has been
studied by using a GEANT3 based Monte Carlo simulation~\cite{wcc0}.
In this paper we report a new study taking into account the water
tank properties such as attenuation length, light collection
efficiency and its angular dependence, etc, based on prototype
measurements. The optical process inside the water tank is
parameterized by an efficiency distribution, $\epsilon(\theta,x)$,
where $\theta$ is the incident angle of charged particles and x the
position of energy deposit. Such a parameterization is obtained by a
GEANT4 based Monte Carlo simulation as described in section 2.2, and
confirmed by the prototype measurement as shown in Fig.\ref{angle}.

\begin{figure}[htbp]
\begin{center}
\includegraphics[width=0.65\textwidth]{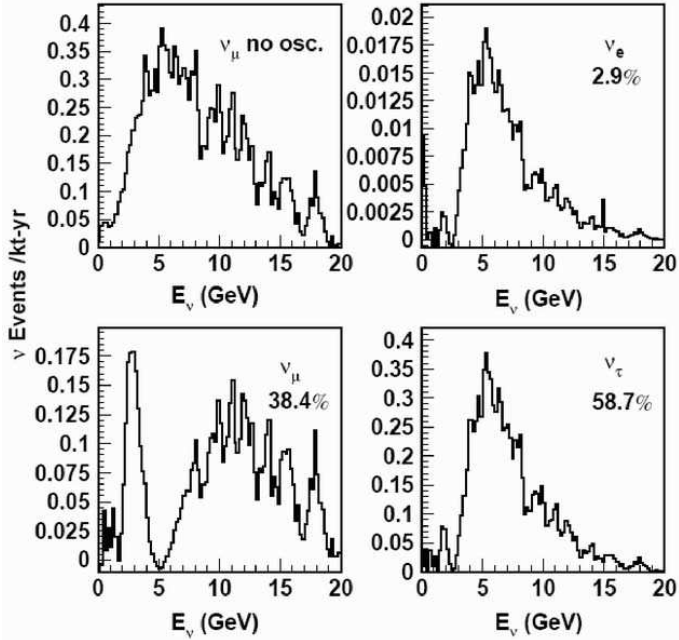}\\
\begin{minipage}[c]{10cm}
\caption{Beam profile of JHF-Beijing with oscillations.
\label{eneu}}
\end{minipage}
\end{center}
\end{figure}

\begin{figure}[htbp]
\begin{center}
\includegraphics[width=0.75\textwidth]{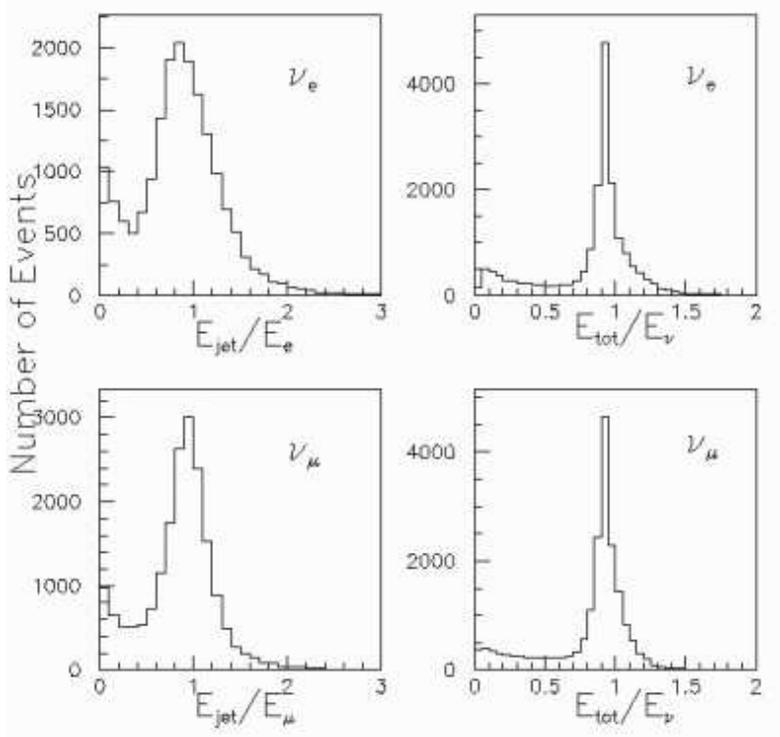}\\
\begin{minipage}[c]{10cm}
\caption{The reconstructed jet energy and the total visible energy.
The fact that $E_{jet}/E_{lepton}$ peaks around one shows that the
jet reconstruction algorithm finds the lepton from CC events. The
fraction of total visible energy to the neutrino energy indicates
that we have an energy resolution better than 13\% for all neutrinos
. The bias is due to invisible neutral hadrons and charged particles
below Cherenkov thresholds. \label{neuresol}}
\end{minipage}
\end{center}
\end{figure}

One possible application considered is the neutrino beam from
JAERI~\cite{japrc} to Beijing with a baseline of
2100km~\cite{chen01}. A total of 60k $\nu_e,\nu_{\mu}$ and
$\nu_{\tau}$ events(Fig.\ref{eneu}) are simulated using an event
generator from the Minos experiment. A $\nu$ CC signal event is
identified by its accompanying lepton, reconstructed as a jet. Fig.
\ref{neuresol} shows the jet energy normalized by the energy of the
lepton. It can be seen from the plot that leptons from CC events can
indeed be identified and the jet reconstruction algorithm works
properly. It is also shown in the figure that the energy resolution
of the neutrino CC events is about 13\% in both cases.

\begin{figure}[p]
\begin{center}
\includegraphics[width=0.65\textwidth]{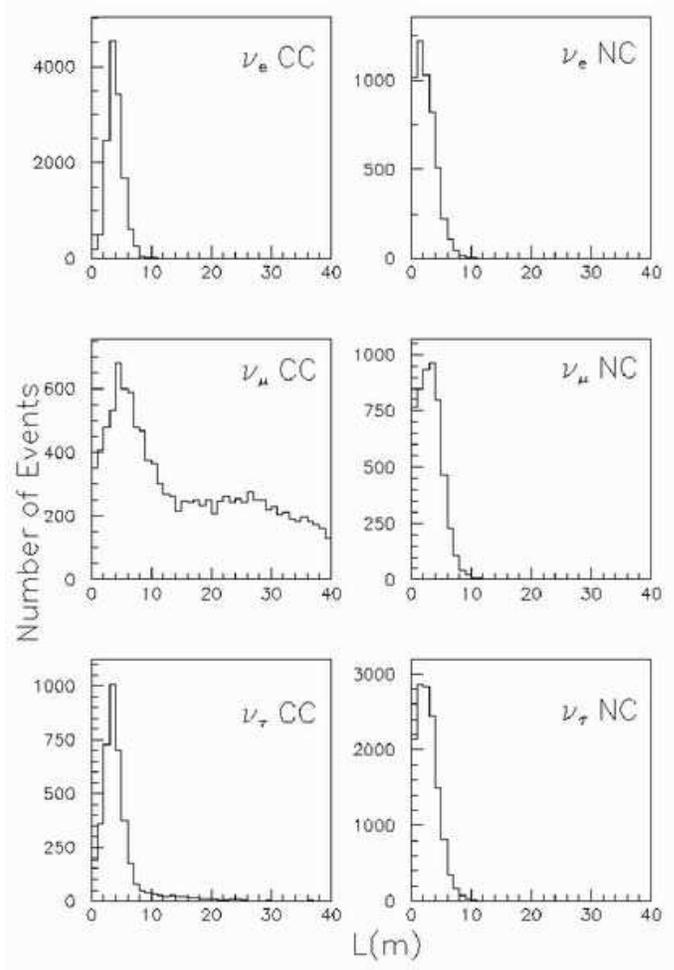}\\
\begin{minipage}[c]{10cm}
\caption{longitudinal length of the jet from shower maximum to the
last cell. \label{lshower}}
\end{minipage}
\end{center}
\end{figure}

\begin{figure}[p]
\begin{center}
\includegraphics[width=0.65\textwidth]{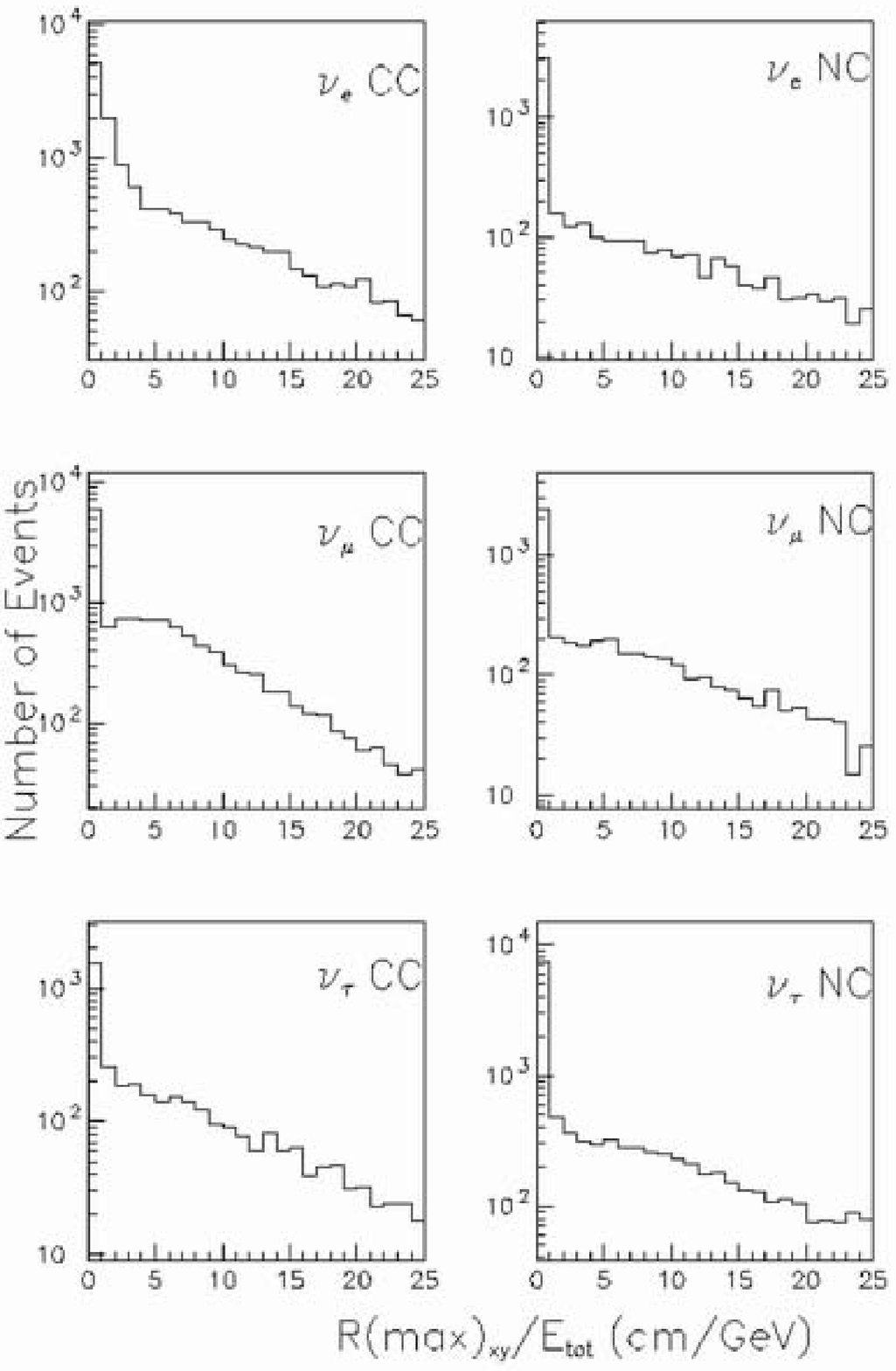}\\
\begin{minipage}[c]{10cm}
\caption{the transverse event size at the shower maximum normalized
to the total visible energy. \label{rmaxetot}}
\end{minipage}
\end{center}
\end{figure}

\begin{figure}[p]
\begin{center}
\includegraphics[width=0.65\textwidth]{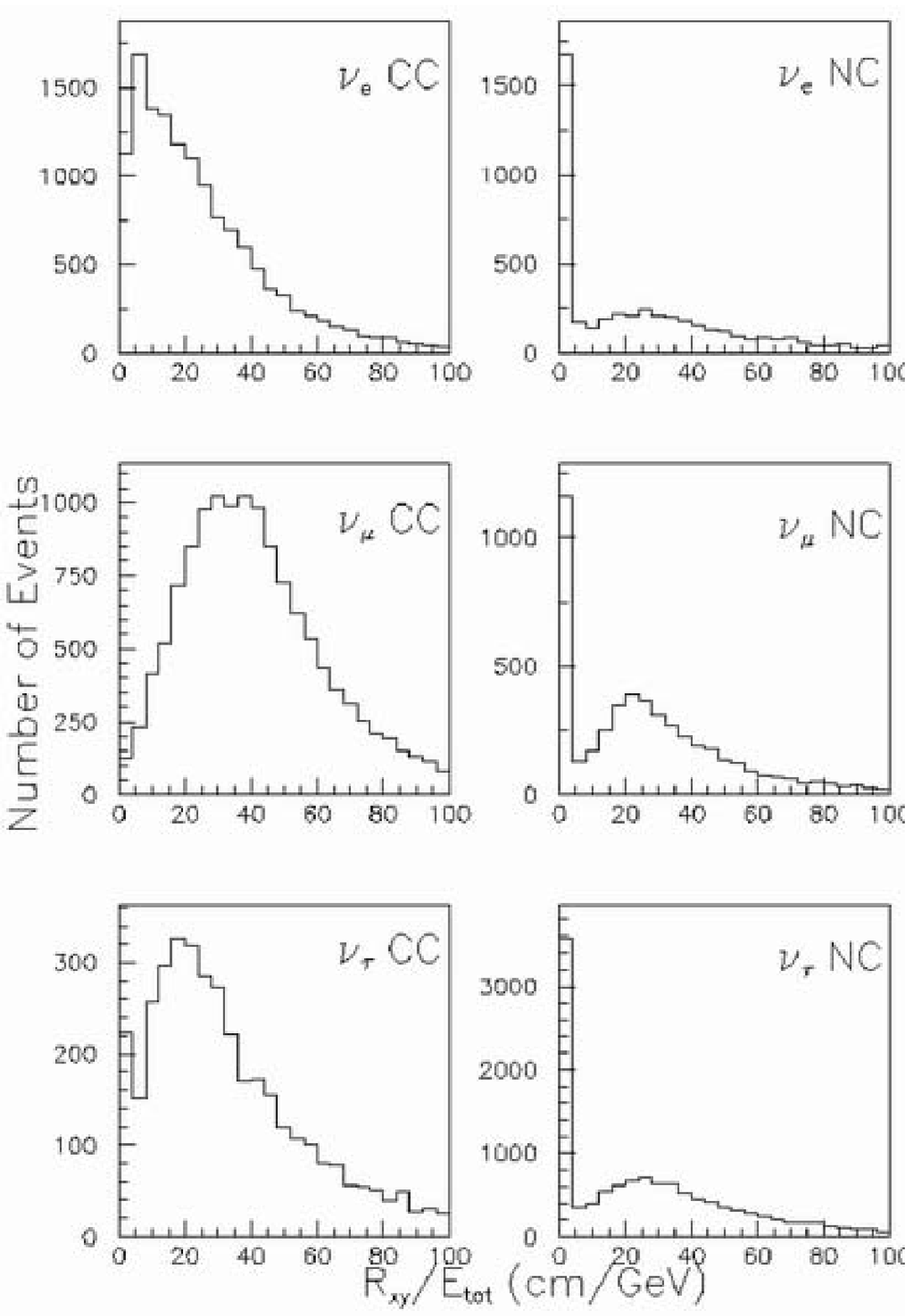}\\
\begin{minipage}[c]{10cm}
\caption{the transverse event size normalized to the total visible
energy. \label{rxyetot}}
\end{minipage}
\end{center}
\end{figure}

\begin{figure}[p]
\begin{center}
\includegraphics[width=0.65\textwidth]{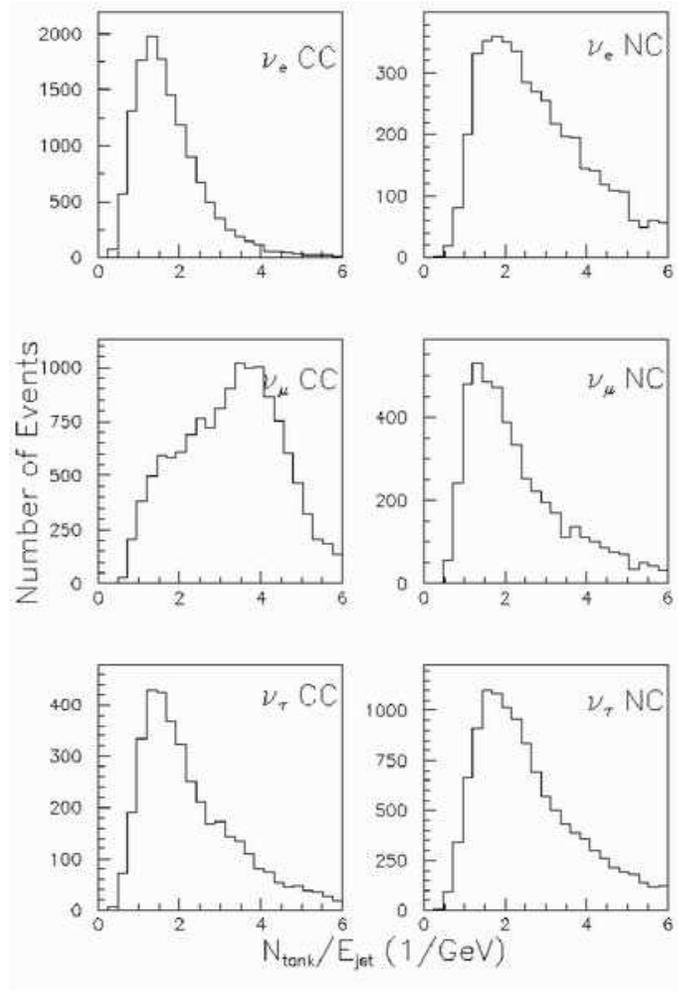}\\
\begin{minipage}[c]{10cm}
\caption{the number of cells with energy above the energy threshold
normalized to jet energy. \label{ntankejet}}
\end{minipage}
\end{center}
\end{figure}

\begin{figure}[p]
\begin{center}
\includegraphics[width=0.65\textwidth]{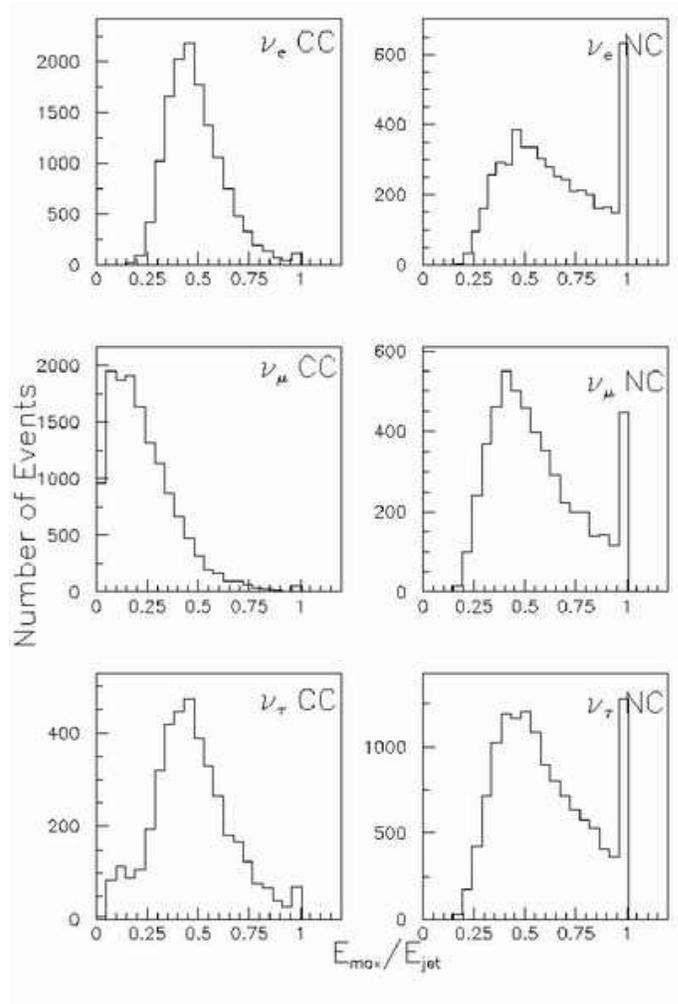}\\
\begin{minipage}[c]{10cm}
\caption{the maximum energy deposition in one cell normalized to jet
energy. \label{emaxejet}}
\end{minipage}
\end{center}
\end{figure}

Since the detector is made of a matrix of water
tanks(Fig.\ref{fig:event}), it's convenient and simple to obtain the
following variables and their distributions for all different
neutrino flavors, which are very effective to identify the neutrino
CC events:
\begin{itemize}
\item L after shower maximum: the longitudinal length of the jet from
the shower maximum to the last cell, as shown in Fig. \ref{lshower}.
Since leptons are only generated by CC process, this varible is good
to distinguish the neutrino CC events and NC events. After the
application of the cut $L>10m$, the remaining $\nu_{\mu}$ NC events
is only 0.1\%;

\item $R_{max}/E_{tot}$: the transverse event size at the shower
maximum normalized to the total visible energy $E_{tot}$, as shown
in Fig. \ref{rmaxetot}. The difference between CC and NC events is
not too much, but combined with others cuts, it can be used to
effectively remove the backgrounds from each neutrino flavor.

\item $R_{xy}/E_{tot}$: the transverse event size normalized to
$E_{tot}$, as shown in Fig. \ref{rxyetot}. It's useful for the
$\nu_e$ and $\nu_{\tau}$ events. For example, after application of
the cut $R_{xy}/E_{tot}<10.$, the ratio of $\nu_e$ NC to CC events
is about 0.38.

\item $N_{tank}/E_{jet}$: the number of cells with energy above the
 threshold(10MeV) normalized to the jet energy of the event
$E_{jet}$, as shown in Fig. \ref{ntankejet}. For the $\nu_e$ and
$\nu_{\tau}$ events, most of NC events can be removed with certain
range of $N_{tank}/E_{jet}$.

\item $E_{max}/E_{jet}$: the maximum energy deposition in one cell
normalized to $E_{jet}$, as shown in Fig. \ref{emaxejet}. A total of
75.6\% of $\nu_e$ CC events can be obtained after the application of
the cut $0.3<E_{max}/E_{jet}<0.6$.
\end{itemize}

\begin{table}[htbp]
\begin{center}
\begin{minipage}[c]{10cm}
\caption{Results from this Monte Carlo simulation: efficiency vs
 background rejection power for different favors. \label{effbg}}
\end{minipage}
\begin{tabular}{c|c|c|c}\hline
&$\nu_e$&$\nu_{\mu}$&$\nu_{\tau}$ \\ \hline CC Eff.&
22\%&51\%&15\%\\ \hline
$\nu_e$ CC  &-& $>$13720:1&7:1 \\
$\nu_e$ NC  &89:1& $>$6110:1&55:1 \\
$\nu_{\mu}$ CC &386:1&-&497:1 \\
$\nu_{\mu}$ NC &45:1&2999:1&38:1 \\
$\nu_{\tau}$ CC&15:1&18:1&- \\
$\nu_{\tau}$ NC &84:1&15802:1&48:1 \\ \hline
\end{tabular}
\end{center}
\end{table}

\begin{table}[htbp]
\begin{center}
\begin{minipage}[c]{10cm}
\caption{After the application of the cut $E_{\nu}>4GeV$, the
results from this Monte Carlo simulation: efficiency vs
 background rejection power for different favors¡£\label{effbg2}}
\end{minipage}
\begin{tabular}{c|c|c|c}\hline
$E_{\nu}>4GeV$&$\nu_e$&$\nu_{\mu}$&$\nu_{\tau}$ \\ \hline CC Eff.&
33\%&71\%&11\%\\ \hline
$\nu_e$ CC  &-&740:1&6:1 \\
$\nu_e$ NC  &31:1&1033:1&37:1 \\
$\nu_{\mu}$ CC &59:1&-&66:1 \\
$\nu_{\mu}$ NC &14:1&434::1&14:1 \\
$\nu_{\tau}$ CC&9:1&13:1&- \\
$\nu_{\tau}$ NC &300:1&593:1&350:1 \\ \hline
\end{tabular}
\end{center}
\end{table}

Table\ref{effbg} shows the final results from this Monte Carlo
study. It can be seen that $\nu_e$ CC events can be selected with
reasonable efficiency and moderate backgrounds. For $\nu_e$ and
$\nu_{\mu}$ events, $\nu_{\tau}$ CC events are dominant backgrounds,
while for $\nu_{\tau}$ , the main background is $\nu_e$. It is
interesting to see that this detector can identify $\nu_{\tau}$ in a
statistical way. With the application of the cut $E_{\nu}>4GeV$,
much higher detection efficiencies of CC events can be obtained, the
results are shown in the Table\ref{effbg2}.  These results are
similar to or better than those from water Cherenkov image detectors
and iron calorimeters~\cite{dick}. We would like to point out that
using sophisticated jet reconstruction algorithms, shower shape
analysis and neural network technique, better results are expected.
Using Table\ref{effbg}, we can explore the sensitivity of our
detector to various quantities, such as $\sin^22\theta_{13}$, CP
phase,etc. Assuming that the systematic error is not dominant, a
sensitivity of 0.2\% to $\sin^22\theta_{13}$ at relevant $\Delta
m^2_{32}$ can be reached for 500kt$\cdot$yr with the designed beam's
intensity of JAERI to Beijing.

\section{Summary}

A full size water tank prototype, with a dimension of $1 \times 1
\times 13m^3$, made of PVC with reflective inner liner was built.
The effective attenuation length and the angular response of the
tank was measured, and good agreement with a GEANT4 based full Monte
Carlo simulation was obtained. The light yield, the total light
collection efficiency, the effective attenuation length and the
angular dependent response of the tank are all good enough for the
long baseline neutrino oscillation experiment, and can be further
improved.  The performance is excellent for $\nu_e$ and $\nu_{\tau}$
appearance and $\nu_{\mu}$ disappearance from the GEANT3 based Monte
Carlo simulation. The water Cherenkov calorimeter is a cheap and
effective detector for $\nu$ factories and super $\nu$ beams, and
such a detector is also desirable for cosmic-ray physics and
astrophysics.There are no major technical difficulties although
further R\&D and detector optimization are needed.

\section{Acknowledgement}
This work is supported by the National Natural Science Foundation of
China under contract No. 10225524 and the Chinese Academy of Science
under contracts No. U-18 and U-35. We would like to thank X.-C. Meng
for his technical assistance to the work presented here. We also
acknowledge the fruitful discussions with Prof. C.-G. Yang and Prof.
J. Cao.






\end{document}